# 1/f and Random Telegraph Noise of Single-Layer Graphene Devices with Interdigitated Electrodes

Georgia Samara
Institute of "Nanoscience and Nanotechnology"
NCSR "Demokritos"
Ag. Paraskevi, Greece
g.samara@inn.demokritos.gr

Nikolaos Vasileiadis
Institute of Nanoscience and Nanotechnology
NCSR "Demokritos"
Ag. Paraskevi, Greece
n.vasiliadis@inn.demokritos.gr

Alexandros Mavropoulis
Institute of Nanoscience and Nanotechnology
NCSR "Demokritos"
Ag. Paraskevi, Greece
a.mavropoulis@inn.demokritos.gr

Christoforos Theodorou
Univ. Grenoble Alpes
Univ. Savoie Mont Blanc
CNRS, Grenoble INP, IMEP-LAHC
Grenoble, France
christoforos.theodorou@grenoble-inp.fr

Konstantinos Papagelis
Department of Physics
Aristotle University
of Thessaloniki
Thessaloniki, Greece
kpapag@physics.auth.gr

Panagiotis Dimitrakis
Institute of Nanoscience and Nanotechnology
NCSR "Demokritos"
Ag. Paraskevi, Greece
p.dimitrakis@inn.demokritos.gr

*Abstract*—**Single-layer Graphene (SLG) is a promising material for sensing applications. High performance graphene sensors can be achieved when Interdigitated Electrodes (IDE) are used. In this research work, we fabricated SLG micro-ribbon (GMR) devices with IDE having different geometric parameters. 1/f noise behavior was observed in all of the examined devices, and in some cases random telegraph noise (RTN) signals suggesting that carrier trapping/de-trapping is taking place. Our experimental results indicate that the geometrical characteristics can have a crucial impact on device performance, due to the direct area dependence of the noise level.**

## I. INTRODUCTION

Although single layer graphene has the thickness of an atom, this 2D material due to its π-conjugation leads to remarkable mechanical, thermal, and electrical properties[1]. Graphene is an emerging material for *More than Moore* electronic devices. Among them, sensors are of special importance especially for wearable and IoT applications. Furthermore, IDE topology has been adopted in capacitive or impedimetric sensor technologies. Parameters such as device structure, sensitivity, fabrication costs and time response play an important role for device commercialization [2]. Low frequency noise (LFN, <100 kHz) in electronic devices provides significant information about the device performance and/or the quality of the fabrication process (e.g., material properties, contacts) [3]. Generation-recombination centers, phonon lattice scattering, nature and density of bulk and/or interface traps, impurities and vacancies can be identified through LFN measurements [4]. In 2011, Balandin *et al.* [5] summarized his LFN experimental results on Graphene FETs. Nah *et al.* [4], studied LFN on graphene with Interdigitated Electrodes (IDE) structures. In this work, we present the effect of GMR and IDE geometrical characteristics on LFN and RTN signals. The correlation between the geometrical parameters and electrical characterization results is expected. This will allow us to optimize the signal-to-noise-ratio (SNR) of the future sensor devices.

## II. EXPERIMENTAL

*A. Device Fabrication*

SLG was grown by chemical vapor deposition (CVD) technique on copper foil. Following, SLG was covered with Poly (methyl methacrylate) (PMMA). After Cu etching, the SLG/PMMA was transferred by wet process onto 300 nm thermally grown $SiO_2$ on Si substrate. UV exposure (325nm) followed by Methyl Isobutyl Ketone (MIBK)/Isopropanol (IPA) cleaning steps were applied to remove PMMA residuals. Finally, hydrogen annealing was followed to improve SLG properties. GMR of different widths (W) 50, 100 and 200μm, were patterned by electron beam lithography (EBL) and Oxygen Plasma (dry) etching. The length of GMRs was 1000μm. Finally, e-beam resist was removed with standard acetone/IPA cleaning steps (Fig. 1). Furthermore, thermally evaporated Aluminium IDE were formed by EBL and metal lift-off technique using. Three different IDE geometries were tested having 8, 15 and

25μm distance (G) among the fingers (Table 1). In total, nine devices were fabricated and tested with various electrical characterization techniques.

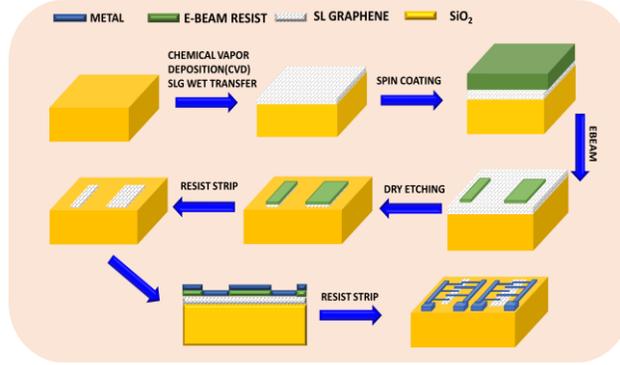

Fig. 1. Experimental steps for the fabrication of IDE devices

TABLE I. TABLE CHARACTERISTICS OF THE EXAMINED DEVICES

| Device Name | Gap (μm) | GMR width (μm) |
|---|---|---|
| G8W50 | 8 | 50 |
| G8W100 | 8 | 100 |
| G8W200 | 8 | 200 |
| G15W50 | 15 | 50 |
| G15W100 | 15 | 100 |
| G15W200 | 15 | 200 |
| G25W50 | 25 | 50 |
| G25W100 | 25 | 100 |
| G25W200 | 25 | 200 |

*B. Electrical Characterization Techniques*

Current –Voltage (I-V) measurements were performed with a HP4155A Semiconductor Parameter Analyzer. LFN/RTN measurements obtained on a wafer prober utilizing Tektronix 4200A (voltage source), I/V converter (SR570) and a digital oscilloscope (DSO7104A) were used (Fig. 2).

III. RESULTS AND DISCUSSSION

*A. Current -Voltage Characteristics*

Initially, all fabricated devices (experimental section) were characterized by *I-V* measurements without any backside bias (Fig. 3). The variable factor for our measurements was the current within the range 1-100μA. The current does not exceed 0.1mA in order to avoid self-heating that destroys *I-V* linearity and protect graphene from oxidation in room ambient.

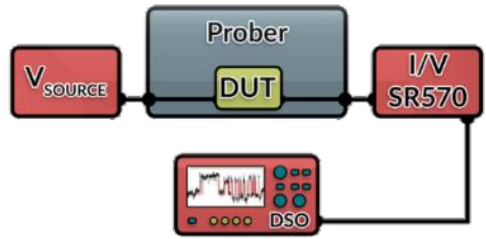

Fig. 2. Setup used for Noise measurements

The samples that were first characterized by *I-V* measurements and shown in Fig. 3 have a fixed channel length of 8μm but different SLG widths of 50, 100 and 200 μm (black, red and blue symbols, respectively). The aim of these electrical measurements was to verify the effect of the channel width of GMR. In Fig.3, ohmic conduction is verified by the linear fitting $ln(V)=ln(R) + n \cdot ln(I)$ results to $n$ in the range 0.98 to 1. This suggest that the work function difference of graphene and Al is small demonstrating the Ohmic behavior of the devices. According to our experimental results, above the current value of 20μA self-heating effect starts taking place. Especially, for the narrowest GMR, the experimental data start deviating significantly from the linear fitting above 20 μA indicating the presence of self-heating effects. Additionally, Meng et al. [6] and Islam et. al. [7], claimed that self-heating effect do exist in GMR with channel width in the range of micro-scale. In our experiments (Fig.3), this effect is diminished for higher channel widths (100 and 200μm). Consequently, ohmic behavior and its direct dependence on GMR width was approved. However, for more in-depth knowledge about IDE GMR devices performance noise measurements was performed. 1/f noise offers important information regarding defects and scattering mechanisms, as well as electronic transport and screening mechanisms in two-dimensional materials [8].

*B. Noise Measurements*

The setup depicted in Fig. 2 has been used for LFN measurements ($f$ <100 kHz) with voltage bias range at $0.1V \leq V \leq 0.6V$. The same triplet of devices was studied in detail. Current recordings for 100 s under constant bias were performed. By applying Fast Fourier Transform on the recorded time series, the Power Spectral Density (PSD) for each measurement was calculated. Fig. 4(a) is a typical example of PSD spectra, for the G8W200 device. A 1/$f$ noise behavior is observed for all gate voltages. All devices were recorded under the same conditions and in all of them 1/$f$ behavior was observed. Here, it should be stressed that the examined GMR devices with IDE can be considered as ($N_{finger}$-1) bottom gate GFETs connected *in series* (S-D-S-D-S-…), where the right and left-side fingers act as Source and Drain electrodes respectively. Also, all measurements performed with floating bottom gate. Such a consideration allows us to apply the standard MOSFET LFN noise theory in order to explain the behavior of the examined devices.

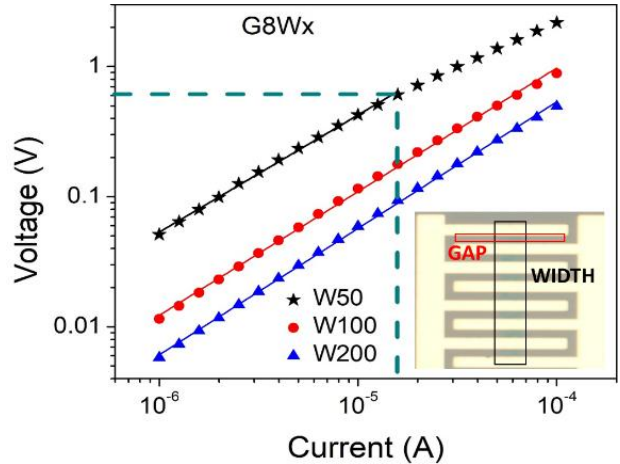

Fig. 3. I-V characteristics of devices with ribbon widths 50, 100 and 200μm and fingers' distance 8μm. The solid lines suggest ohmic region. Inset: Optical microscope image of the presenting the metal fingers and the GMR (dashed area) of G8W50 examined device.

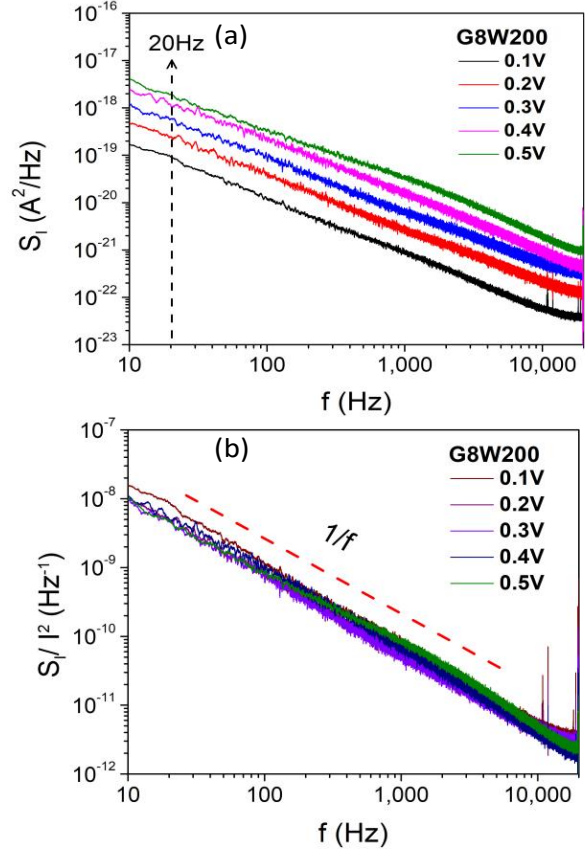

Fig. 4. Typical (a) PSD $S_I$ and (b) normalized PSD $S_I/I^2$ under different constant bias voltages of the examined device with gap (finger inter-distance) 8μm and GMR width 200μm.

Furthermore, the normalized PSD, $S_I/I^2$, is plotted in Fig.4(b) vs frequency $f$ indicating ~$1/f^\gamma$ ($0.8 < \gamma < 1.2$) dependence. This behavior has been observed by Balandin et. al. [5], but not on devices IDE geometry. Also, we can clearly observe that from 10 Hz up to 10 kHz, the PSD spectra almost coincide, indicating a coherence of noise behavior across all applied gate bias.

Based on these experimental data, the next step was to focus on PSD values at a chosen frequency of 20 Hz in order to study the noise dependence with voltage or current and gain an insight about the origin of $1/f$ noise. For this reason, in Fig. 5, all PSD values at 20 Hz were collected for devices with constant gap and variable channel width, and $S_I^{1/2}$ values were calculated and plotted for different applied voltages. It is important to mention that there is a direct correlation of $1/f$ amplitude and channel width (GMR width), indicating that the measured noise is not generated by the instrumentation. Along with this observation, there is also an increasing dependence on the applied voltage, which is in fact linear, as demonstrated by the linear regression. Thus, we can extract the equivalent resistance noise, $S_R$, from the respective slope for each geometry.

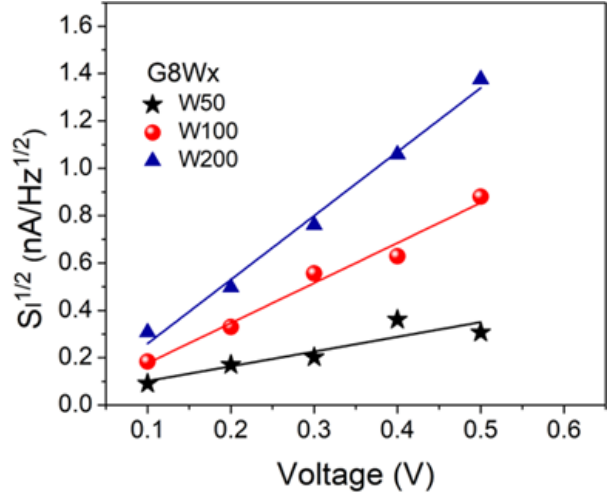

Fig. 5. Plots of the square root of PSD at 20Hz vs voltage for devices with different GMR widths and the same fingers distance (8μm). Lines denote linear fit to experimental data.

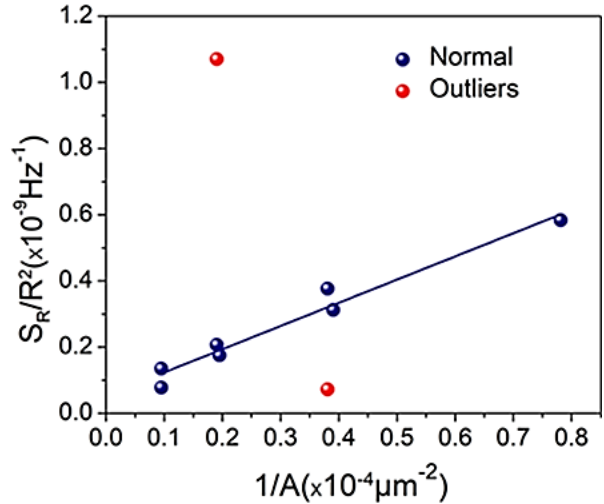

Fig. 6. In total nine devices with different ribbon widths (50, 100, 200 μm) and gaps (8, 15, 25 μm) were examined. Line indicates the linear fit to normal data.

A total of nine devices was studied (Table 1). Specifically, $S_R$ was extracted for all combinations of three different GMR widths (W = 50, 100, 200 μm) and three different gaps (G = 8, 15, 25 μm). Fig. 6 illustrates the dependence of the normalized resistance noise, $S_R/R^2$, versus $1/A$, where $A = G_{gap} \times W_{width} \times (N_{fingers}-1)$ is the graphene active area. A clear correlation is observed between the noise and the geometrical characteristics of the samples, for all gaps and widths, except G15W50 and G25W100 devices. The latter are called outliers, due to their deviation from the linear fitting (Fig.6), which may be attributed to graphene grain boundaries, wrinkles or to the edges of the GMR (armchair, zig-zag) [9].

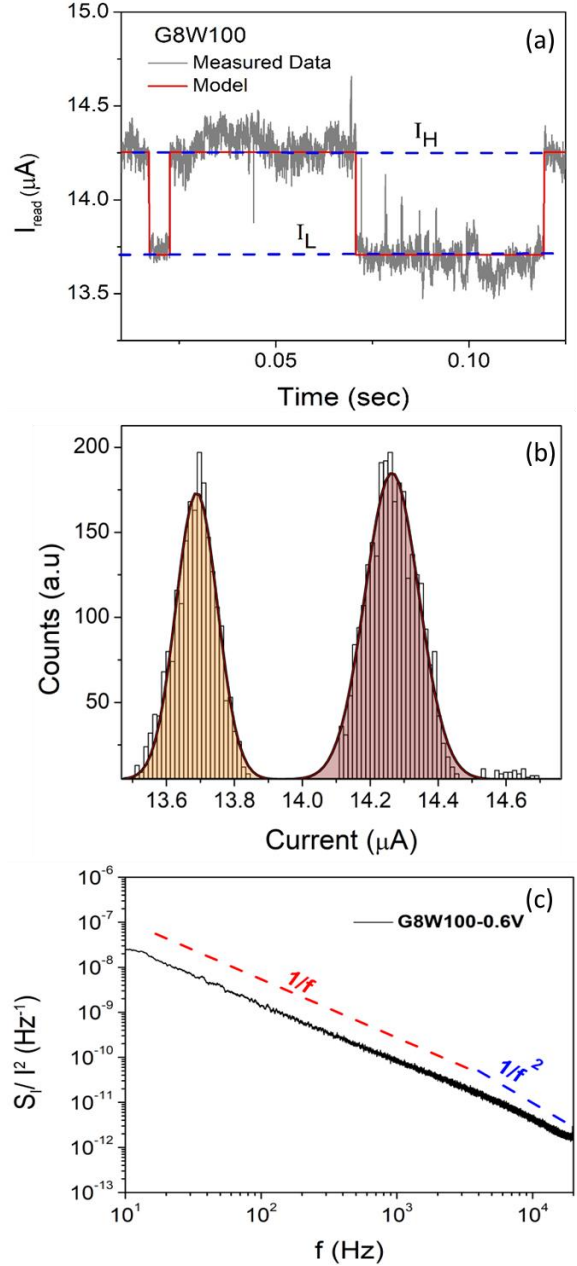

Fig. 7. Typical (a) RTN signal with two current levels, (b) histogram of the current values with Gaussian distribution fitting and (c) normalized PSD spectra, for G8W100 device measured at 0.6V.

It can be concluded without doubt that the higher the active graphene area, the higher the SNR. Therefore, the experimental data of Fig. 5 suggest that LFN does not originate only from the interface of graphene with Al contacts. Consequently, Fig. 5 and 6 suggest the resistive nature of the measured noise, meaning it is due to conductance fluctuations concerning all the active graphene area and not only the contact regions.

*C. Random Telegraph Noise*

Besides PSD analysis, *I-t* recordings were also examined extensively for Random Telegraph Noise (RTN), which was detected for a few devices amongst the one is presented in Fig. 6. Fig.7 (a) presents measured data (gray line) for a typical two-level RTN from G8W100 device at 0.6V. This figure is not showing the total RTN signal, but refers to a restricted and short time domain. Moreover, after statistical analysis, to extract the pulse durations, the resulting pulse fit is denoted with the continuous (red) line. The two (blue) dotted lines correspond to the High /Low levels of the signal, where $I_H$ and $I_L$ denote the high and the low current level values. According to the classical theory of RTN, $I_H$ and $I_L$ levels are due to electron capture and emission respectively from a trap. In our case, these trapping sites are lying at the interface between SLG and the $SiO_2$ (substrate) where the SLG was transferred [10]. Fig. 7(b) presents the corresponding histograms of the current signal values. Obviously, the two main histograms of current values, related to $I_H$ and the $I_L$, respectively. Excellent histogram fitting (black line) with Gaussian distributions is achieved. Fig. 7(c) is the PSD spectra pertaining to the examined device, wherein theoretical $1/f$ and $1/f^2$ behaviors are delineated by the red and blue dashed lines, respectively.

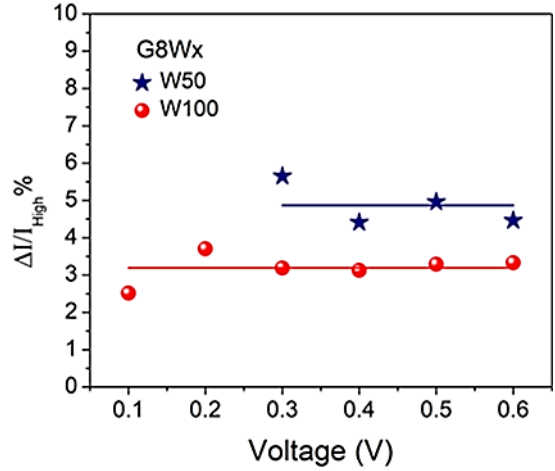

Fig. 8. Graphs of $\Delta I/I_H$ vs Voltage for the same gap (8μm) and GMR widths 50 and 100 μm. Lines denote the mean value of the data clusters.

In addition, RTN was studied on different geometries as shown in Fig.8. The reasoning for Fig.8, was also to understand the effect of the width of GMR on RTN amplitude. Devices with fixed gap of 8μm and GMR width 50 and 100, were investigated. No RTN signals were found on the G8W200 device. The first step was to distinguish for which applied voltages RTN was present and to calculate the factor $\Delta I/I_H$ (%) for each signal with RTN, where $\Delta I$ is the mean value of all the differences between $I_H$ and $I_L$, for all the RTN at each voltage bias. The extracted $\Delta I/I_H$ is plotted in Fig. 8 versus the applied voltage. Also, it is important to note that the escalation of the voltage from low to high values, has almost no impact on the $\Delta I/I_H$ values for the examined devices, i.e., $\Delta I/I_H$ remains almost constant with bias voltage. The solid lines in Fig. 8 correspond to the mean value of $\Delta I/I_H$ across all voltages for each device.

Furthermore, correlation between the GMR geometry and RTN amplitude is observed: lowering the GMR width, the noise level increases, exactly as in the case of $1/f$ noise. Therefore, in conjunction with Fig. 6, RTN trapping can be assumed to originate from interface states between the GMR and the $SiO_2$ substrate.

## IV. CONCLUSIONS

Static characteristics of GMR with IDE reveal direct dependence of the device with channel width and validate Ohmic conduction up to 20 μA currents, where self-heating effects start taking place for narrow GMR. In addition, the current noise has a $1/f$ behavior for $0.1\ V \leq V \leq 0.6\ V$, indicating a defect uniformity. It was attributed to conductance fluctuations, most probably originating in the active graphene area, due to the ($1/area$)-scaling of the noise level, and not the access regions. Finally, RTN was found to have the same dependence with area as $1/f$ noise, suggesting that trapping /de-trapping events take place between GMR and surface defects of $SiO_2$.

## ACKNOWLEDGMENT

This work was supported in part by the research projects "3DTOPOS" (MIS 5131411) and "LIMA-chip" (Proj. No. 2748) which are funded by the Operational Programme NSRF 2014-2020 and the Hellenic Foundation of Research and Innovation (HFRI) respectively.